\documentclass[onecolumn,prd,superscriptaddress,showpacs,floatfix,%  
preprintnumbers,nofootinbib,showkeys]{revtex4-1}  

\usepackage[dvips]{color}
\usepackage[normalem]{ulem}
\definecolor{darkgreen}{cmyk}{1,0,1,0.4}
\definecolor{darkred}{cmyk}{0,1,1,0.4}

\usepackage{amsmath}
\usepackage{amssymb}
\usepackage{graphicx}
\usepackage{epsfig,epsf}
\usepackage{color}
\def\beq{\begin{equation}}
\def\eeq{\end{equation}}
\def\barr{\begin{array}}
\def\earr{\end{array}}
\def\beqarr{\begin{eqnarray}}
\def\eeqarr{\end{eqnarray}}
\def\dis{\displaystyle}
\def\vt{ V}
\def\mpl{M_{\rm Pl}}
\def\lapp{\mathrel{\rlap{\raise.5ex\hbox{$<$}}
                    {\lower.5ex\hbox{$\sim$}}}}
\def\gapp{\mathrel{\rlap{\raise.5ex\hbox{$>$}}
                    {\lower.5ex\hbox{$\sim$}}}}
\def\yt{\tilde y}
\def\bfig{\begin{figure}}
\def\efig{\end{figure}}
\renewcommand{\l}{\left}
\renewcommand{\r}{\right}
\newcommand{\nn}{\nonumber}
\newcommand{\bib}{\begin{thebibliography}}
\newcommand{\ebib}{\end{thebibliography}}
\begin{document}
\title{ A Geometric Approach to Modulus Stabilization}

\author{Sampurn Anand}
\email{jha.sampurna@gmail.com}
\author{Debajyoti Choudhury}
\email{debajyoti.choudhury@gmail.com}
\affiliation{Department of Physics and Astrophysics, 
University of Delhi, Delhi 110007, India}
\author{Anjan A. Sen}
\email{aasen@jmi.ac.in}
\affiliation{Centre for Theoretical Physics, Jamia Millia Islamia
New Delhi 110025, India.}
\author{Soumitra SenGupta}
\email{tpssg@iacs.res.in}
\affiliation{Department of Theoretical Physics, Indian Association for the 
Cultivation of Science, % 2A \& 2B, Raja S.C. Mullick Road, 
Kolkata 700032, India.}

\date{\today}
\pacs{}
\keywords{}

\begin{abstract}
Modulus stabilization, a must for explaining the hierarchy problem in
the context of the Randall-Sundrum (RS) scenario, is traditionally
achieved through the introduction of an extra field with {\em ad hoc}
couplings.  We point out that the stabilization can, instead, be
achieved in a purely geometrodynamical way, with plausible quantum
corrections in the gravity sector playing the key role.  The size of
the corrections that lead to acceptable phenomenology is also
delineated.
\end{abstract}
\maketitle

%\section{Introduction}
Notwithstanding the recent discovery of the Higgs
boson\cite{cms-higgs,atlas-higgs}, the lack, so far, of any definitive
signature of physics beyond the Standard Model (SM) is
perplexing. Although the mass of the Higgs boson is such that new
physics at a nearby scale is not demanded by considerations of
triviality or vacuum stability, turning this around to imply that none
exists until the Planck scale ($\mpl$) is, at the least, aesthetically
repugnant. Indeed, the hierarchy problem of the SM continues to be a
vexing issue, and, over the years, several mechanisms have been
suggested to ameliorate this. While most of these scenarios also do
promise explanations of some of the other puzzles that beset the SM,
no direct evidence for any of the new states intrinsic to these
theories have been seen so far. Furthermore, several of these have,
associated with them, some form of a little hierarchy problem.

An interesting approach (RS) to the hierarchy problem essentially does
away with a fundamental weak scale, ascribing the apparent hierarchy
to a geometrical origin~\cite{rs}. Envisaging space-time to be a slice
of ${\rm AdS}_5$, the known world is confined to one of a pair of
three-branes that sit atop the two fixed points of a $S^{1}/Z_{2}$
orbifold. The metric has the non-factorizable form
%%%%
\beq
ds^{2} = e^{-2kr_{c}|y|}\eta_{\mu\nu}dx^{\mu}dx^{\nu} + r_c^2 \, dy^{2} \ ,
\label{line_element}
\eeq
%%% 
with $\eta_{\mu\nu} = diag(-1,1,1,1)$ being the Minkowski metric and
$y\in[0,\pi]$ with $(x^{\mu}, -y) \equiv (x^{\mu}, y)$.  On the
(visible) brane at $y=\pi$, the natural mass scale is suppressed by a
factor of $e^{-kr_{c}\pi}$ with respect to the fundamental scale, {\em
e.g.} that operative at the (hidden) brane located at $y =0$. With
$k\approx {\cal O}(M)$ arising naturally, having $kr_{c}\approx 12$
would `solve' the hierarchy problem. However, the modulus $r_{c}$ is
not determined by the dynamics. This can be cured by promoting $r_{c}$
to a dynamical field ${\cal M}$ and inventing a mechanism that forces
it to settle to ${\cal M} = r_{c}$.

To this end, ref.\cite{gw} introduced a new scalar field $\phi$ in the
bulk with a quadratic potential. Interacting, as it does, with ${\cal
M}$ through the metric, integrating out $\phi$ would result in an
effective potential $V_{\rm eff}({\cal M})$.  A suitable form for
$V_{\rm eff}({\cal M})$ can be arranged if $m_{\phi}\ll M_{\rm Pl}$ as
well as if $\phi$ has brane localized potentials that ensures
appropriate classical value on the branes, leading to 
%%%%%%%%
\beq kr_{c}\simeq
\frac{k^{2}}{m^{2}_{\phi}} {\rm ln}\l(\frac{\phi(y=0)}{\phi(y=\pi)}\r) .
\eeq 
%%%%%%%% 
The apparent success of this (GW) mechanism~\cite{gw} hinges on the {\em
  ad hoc} introduction of a new fundamental scalar, with masses and
couplings being just so. This criticism can also be levelled at
variations of this mechanism that have been
attempted~\cite{Cline:2000xn}. It would be nice if the stabilization
process could have emerged more naturally. To this end, we appeal to a
geometric origin in the shape of corrections to the Einstein-Hilbert
action itself. While this may seem an {\em ad hoc} measure as well,
such corrections are liable to arise in any quantum theory of
gravity. In the absence of a definitive theory, though, we are unable
to determine the exact structure of such corrections and hence
consider the effective action for gravity to constitute all possible
structures consistent with diffeomorphism invariance as well as other
imperatives. This, in general, allows for terms higher order in $R,
R_{ab}R^{ab}$ and $R_{abcd}R^{abcd}$. Inclusion of the last two,
generically, leads to instabilities, although particular linear
combinations escape this fate. On the other hand, the replacement
$R\rightarrow f(R)$ is often free from such
instabilities~\cite{felice,Nojiri:2010wj}.  We start by postulating the
five-dimensional pure gravity action, in the Jordan frame, to be
\beq
\barr{rcl}
S_{EH} & = & \dis \int d^{4}x dy\sqrt{\tilde g}\left(2M^{3}f(\tilde R) - 
          2\lambda M^{5} \right)  \\
&- & \dis \int d^{4}x dy\sqrt{\tilde g}\l[\lambda_{v} \delta(y-\pi) + \lambda_{h}\delta(y)\r], 
\earr
\label{E_H_action}
\eeq 
%%%%%%%%%%%% 
% 
where $M$ is the fundamental mass scale and 
$\tilde g_{ab}$ the metric with $\tilde g = -{\rm
Det}(\tilde g_{ab})$. While it could have been included 
in $f(R)$ itself, we prefer to write the putative 
cosmological term explicitly, with $\lambda \lapp {\cal O}(1)$.
Similarly, $\lambda_{v,h}$ are the 
tensions associated, respectively, with the visible and the hidden brane.

Concentrating on the bulk action, it can be rewritten as
%%%%%%
\beq S_{blk} = \int d^{4}x dy\sqrt{\tilde g}
\l(2M^{3} \tilde R \, F - U - 2\lambda  M^{5}\r),
\label{action_bulk} 
\eeq 
%%%%%
where, 
\beq U =  2M^{3}\l[ \tilde R F- f(\tilde R)\r] \mbox{ and } 
F \equiv f'(\tilde R) \ .  
\eeq 
%%%%% 
The non-minimal coupling above can be rotated away 
by a conformal transformation\cite{weyl, felice,Nojiri:2010wj}, {\it viz.},
%%%%%%
 \beq 
  \tilde g_{ab} \rightarrow g_{ab} 
               = \exp(2 \, \omega(x^{\mu},y)) \, \tilde g_{ab}
\eeq 
%%%%%
with the actual form of $\omega(x^{\mu},y)$ yet to be specified. The Ricci 
scalars in the two frames are related through
%%%%%%%%
\[
 \tilde R = e^{2 \, \omega} \left[R + 8 \, \Box\omega -12
    g^{ab}\partial_{a}\omega\partial_{b}\omega \right] \ ,  
\]
%%%%%%%
with $\Box$ representing the Laplacian operator appropriate for the 
Einstein frame (defined in terms of $g_{ab}$). Choosing 
a specific form of $\omega(x^\mu, y)$, {\it viz.},
%%%%%%%%
\beq
\omega = \frac{1}{3} \, \ln F \equiv \frac{\gamma \, \phi}{5} 
\ , \qquad \gamma \equiv \frac{5}{4 \, \sqrt{3} \, M^{3/2}} \ ,
\eeq
we have,
%%%%
\beqarr
S & = & \dis \int d^{4}x d\yt \, \sqrt{g} 
    \l[2M^{3}R -\frac{1}{2} g^{ab}\partial_{a}\phi\partial_{b}\phi 
  -\vt(\phi)\r] 
 \\
& - & \dis  \int d^4x d\yt \,\sqrt{g} 
      e^{-\gamma \phi} 
      \l[\lambda_{v}\delta(\yt-l) + \lambda_{h}\delta(\yt)\r] 
\eeqarr
\label{action_Eframe}
%%%%%
where
\beq
\vt(\phi) = \l[U(\phi)+2\lambda M^{5}\r] \,
         \exp\l( - \gamma \, \phi \r) \ .
\eeq

We have, thus, successfully traded the complex form of $f(\tilde R)$ for the
usual Einstein-Hilbert action, supplemented by a scalar field that
essentially encapsulates the extra degree of freedom encoded in the
higher powers of derivatives in $f(\tilde R)$. As long as the potential
$\vt(\phi)$ is bounded from below, the system would be free from
Ostrogradski instabilities.

The exact form of $\vt(\phi)$ would, of course, hinge on the form of
$f(\tilde R)$. Some features of the scenario, though, are ubiquitous. Unlike
in the original RS scheme, the two brane tensions would, in general,
be unequal in magnitude. This could have been anticipated in the
Jordan frame as well, for the tensions were necessary to allow for the
discontinuity in the logarithmic derivative of the metric at the
orbifold fixed points; and for $f(\tilde R)\neq \tilde R$, the two junction
conditions can not be expected to be equivalent\footnote{This would
also have been forced upon us in the GW case were back reaction to be
taken into account.}.  What may seem even more problematic is the
existence of the bulk scalar field as we would now need to consider
the coupled system ($g_{ab}, \phi$) instead of the vacuum equations as
examined in ref.\cite{rs}. This can be done though, but usually
results in a very complicated set of equations, which rarely is
amenable to closed form analytic
solutions~\cite{back_reaction}. Furthermore, there are several subtle
issues that invalidate some of the approaches. Rather than follow this
path, and present a set of dense expressions and/or numerical
solutions, we % take recourse to an approximation that turns out to be
% quite an excellent one.
first appeal to a physically 
well-motivated approximation.

Consider the case where $\vt(\phi)$ has a minimum at $\phi=\phi_{\rm
min}$. Given sufficient time, one would expect that $\phi$ would
settle at $\phi_{min}$ with $\vt(\phi_{min})$ acting as the effective
cosmological constant ({\em i.e}, it would assume the role of
$\Lambda$ in \cite{rs}).  To the leading order, only small deviations
about $\phi_{min}$ should need be considered. If the mass of the
``fluctuation field'' is small, then so is the energy contained in it
and neglecting the corresponding back reaction is 
justifiable and constitutes a very useful first approximation. 

\begin{figure}[!htbp]
{
\vspace*{-8pt}
\hspace*{-30pt}
\epsfxsize=7cm\epsfbox{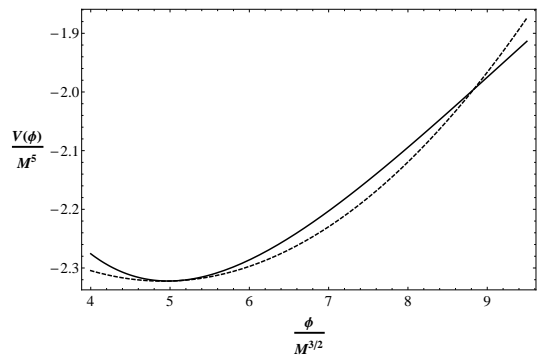}
\vspace*{-8pt}
}
\caption{The solid line denotes the potential for 
$a=0.01, \ b = 0.01, \ 
\lambda =  -0.6$, while the dashed line denotes a 
typical approximation for $\phi_a = 5.05 \, M^{3/2}$.}
\label{fig:potential}
\end{figure}

While much of what follows below can be applied to a wide class of
$f(\tilde R)$, we choose to work with a series expansion 
in $\tilde R/M^2$, retaining only a few terms so as to facilitate
an immediate examination of each step in the analysis, viz. 
%%%%%%%%
\beq
f(\tilde R) = \tilde R + a \, M^{-2} \tilde R^2 + b \, M^{-4} \tilde R^3 . \label{fr}
\eeq
%%%%%%%%
where  $a,b$ are dimensionless 
free parameters with each, presumably, $\lapp {\cal O}(1)$. 
As would be expected, we need $b > 0$ for 
obtaining a sufficiently negative $V(\phi_{min})$ 
as also the desirability of a small second derivative 
at the minimum, whereas $a$ can assume either sign or even vanish. 
It is interesting to note that $f(\tilde R) = \tilde R^\beta$, typically, 
fails the twin test, unless $\beta$ is a certain very specific fraction. 
The corresponding potential has the form 
%%%%
\beq
\vt = 
2 M^{-5} F^{-5/3}\l[ \lambda + a R^2(F) + 2b R^3(F) \r], 
\label{bulk_pot}
\eeq
where (for phenomenological reasons)
we confine ourselves to a specific branch, namely
\beq
R(F) = \frac{-a - \sqrt{a^2-3b(1-F)}}{3b}.
\eeq
%%%%%%%%%%%%%%%%%%%
We look for a situation whereby, in the Einstein frame, 
the only non-trivial dependence of the metric is on the 
coordinate $\yt \equiv r_c \,y$, namely 
\beq
ds^{2} = e^{-2\sigma(\yt)}\eta_{\mu\nu}dx^{\mu}dx^{\nu} + d \yt^{2} 
\label{Eframe_lin_element}
\eeq
The Einstein's equations reduce to
% G55 component
\beq
\barr{rcl}
6\sigma'^2 & = & \dis \frac{1}{4M^3}\l[\frac{1}{2}\phi'^{2}-V\r],
\\[2ex]
3\sigma'' & = & \dis \frac{1}{4M^3}\big[\phi'^{2}+e^{-\gamma\phi}
\l(\lambda_{h}\delta(y)+\lambda_v\delta(y-l)\r)\big] \, 
\label{g00}
\earr
\eeq
whereas the scalar field satisfies,
%EoM for field
\beq
\phi'' - 4\sigma'\phi' - \frac{dV}{d\phi}+\gamma e^{-\gamma\phi}
\l[\lambda_{h}\delta(y)+\lambda_v\delta(y-l)\r]=0.
\label{eom_phi}
\eeq
%%%%%%%%%%%%%%%%%%%%%
Although the system can be solved numerically, it is instructive 
to consider an approximation so as to allow for closed form 
analytic expressions. Expanding around $\phi = \phi_a \sim \phi_{\rm min}$, 
we write 
\beq
\frac{V}{M^5}= V_0 +\l(\frac{ V_1}{M^{7/2}}\r)\xi +
\l(\frac{ V_2}{M^{2}}\r)\xi^2,
\eeq
where $\xi(\yt) = M^{-3/2} \, (\phi-\phi_a)$ and 
$V_i$ are constants. It might seem counterintutitive to consider 
$\phi_a \neq \phi_{\rm min}$; this, however, is useful to enhance 
the applicability of the approximation, which we require to be better 
than $\sim 10\%$ over the range of interest (see Fig.\ref{fig:potential}). 

%%%%%%%%%%%%%%%%%% FIGURE %%%%%%%%%%%
%%%%%%%%

Neglecting the back reaction altogether would reduce the system 
to the standard GW scenario with the corresponding warping
$\sigma_{(0)}(\yt)$ being linear in $|\yt|$ with the coefficient 
determined by $V_0$. However, doing so is not really justified 
(even in the GW case) as it can be as much as 10\% or larger.
Hence, we effect an inclusion by solving the bulk equations iteratively. 
Defining
%%%%
\beq
k^2 \equiv \frac{1}{24M^3}\l[\frac{V_1^2}{4V_2}-V_0\r],
\eeq
%%%%
we introduce the notation
\beq
\kappa_{0,2} \equiv V_1 / V_{0,2} \ , \quad N_\pm \equiv 2 \, \nu \pm 21
\eeq
where $\nu=\sqrt{4+ 2 V_2/ k^2}$. Then, the first order solution 
to the warping is 
%%%%
\beq
\sigma_{(1)} =  k|\yt|+\frac{1}{18M^3}
\l[c_1^2 \, e^{2\alpha_1|\yt|}+
c_2^2 \, e^{2\alpha_2|\yt|} - \, 
\frac{c_1c_2 V_2}{k^2}
e^{4 k |\yt|}\r]
   \label{sigma_1}
\eeq
%%%%
where $\alpha_{1,2} = (2 \pm \nu) \, k$. 
The dimensionless constants
$c_{1,2}$ can be determined by matching the discontinuities, 
leading to ($l \equiv r_c \pi$)
\[
\barr{rcl}
0 & = & \dis N_-^{-1} \, c_{1} -  N_+^{-1} \, c_{2}
+ \l(20\sqrt{3}+ 25 \, \kappa_2\r)\\[1ex]
0 & = & \dis   N_-^{-1} \, c_{1} \, e^{(2+\nu)kl} - 
 N_+^{-1} \, c_{2}e^{(2-\nu)kl} 
    + \l(20\sqrt{3}+ 25 \, \kappa_2\r). 
\earr
\label{eqn_coeff}
\]
For large $kl$ (applicable since we need $kl\approx 36$ to explain the
hierarchy), one obtains
\[
\barr{rcl}
c_{1} & \simeq & \dis
N_{-}^{-1} \, \l(20\sqrt{3} + 25 \kappa_2\r) \, 
\l[ e^{-2\nu kl}-e^{-(2+\nu)kl}\r], \\[1ex]
c_{2} & \simeq & \dis 
N_{+}^{-1} \, \l(20\sqrt{3} + 25 \kappa_2\r) \,
\l[1+ e^{-2\nu kl}-e^{-(2+\nu)kl}\r],  
\earr
\label{sol_coeff}
\]
 
The nonlinear terms in 
eqn.(\ref{sigma_1}) account for the leading backreaction due 
to the scalar field, which, to this order, is given  by
%%%%%%%%
\beq
\xi_{(1)}(\yt) = % \frac{-V_1}{2V_2}
- \kappa_2 / 2
+M^{3/2}\l[c_1e^{\alpha_1|\yt|} 
+ c_2e^{\alpha_2|\yt|}\r]
\label{phi_0}
\eeq
%%%%%%%%
One could extend this to even higher orders, with the additional 
corrections being given in terms of confluent hypergeometric functions. 
Substituting 
eq.(\ref{phi_0}) in the action  and integrate over $\yt$, 
the effective potential for the modulus field is obtained to be
\[
\frac{V_{\rm eff}}{M^3 k}  = 
     \left[ d_{0} + d_{1} (e^{-2\nu kl} - 2 e^{-(2+\nu)kl} )
                       + d_{2}e^{-4kl} \right]
         \]
where
%%%%%%%%%%% d0 %%%%%%%%%%%%%%%
\beqarr
d_0 & = & 24 -\frac{12\kappa_0\kappa_2}{\kappa_0\kappa_2 - 4} \,
 +\frac{(\nu-2)(40\sqrt{3}+25\kappa_2)^2}{4N_{+}^2} \nn \\ 
& & + \frac{25}{16N_{-}^2}\l(5\kappa_2 +8\sqrt{3}\r)\l((67-4\nu)8\sqrt{3}+125\kappa_2\r)  \nn
\\[2ex]
d_{1} & = & \dis 250 \, N_{+}^{-2} \, N_{-}^{-2} \,
           \nu  \l(5 \kappa_2 +8 \sqrt{3}\r) \, 
     \l[\l(\nu ^2+21\r) \, \kappa_2 + 210 \sqrt{3} \r] \nn
\\[2ex]
d_2 & = & \dis \frac{48 \l(21 \nu - \nu ^2 + 46\r)}{N_{-}^2} \, 
   + \frac{250 \sqrt{3} \, (\nu +2) \kappa_2}{N_{-}^2} \nn \\
 & & + \l(\frac{48}{\kappa_0\kappa_2-4} \,
 +\frac{625 (4 \nu -17) \, \kappa_2^2}{16 \, N_{-}^2 }\r) \nn.
 %\earr
\eeqarr
%%%%
The consequent extrema are given by
\beq
e^{(2-\nu)kl} = \frac{(2+\nu)\pm \sqrt{(2+\nu)^2-8\nu d_{2}/d_{1}}}{2\nu }
 \label{roots}
\eeq
%%%%%%%%%%%%%%%%%% FIGURE %%%%%%%%%%%

\begin{figure}[!h]
\begin{center}
\includegraphics[width=2.6in,height=1.5in,angle=0]{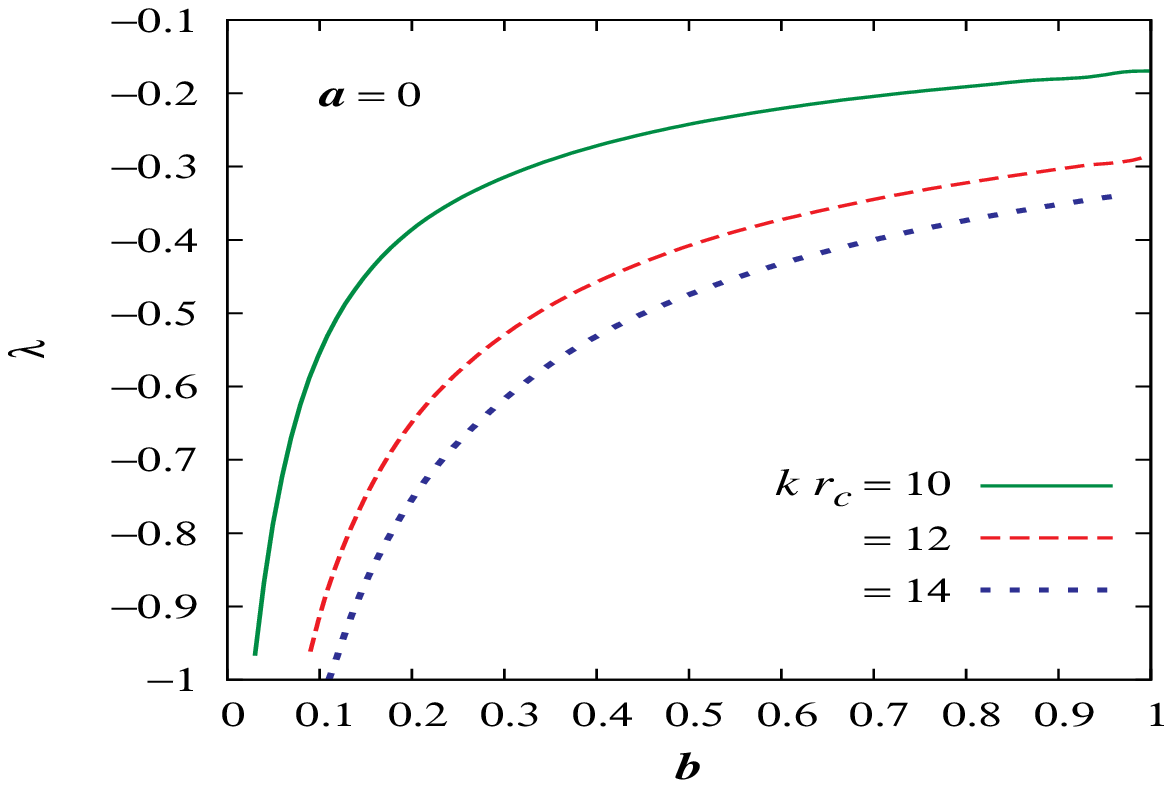}
%\vspace*{-2ex}
\includegraphics[width=2.6in,height=1.5in,angle=0]{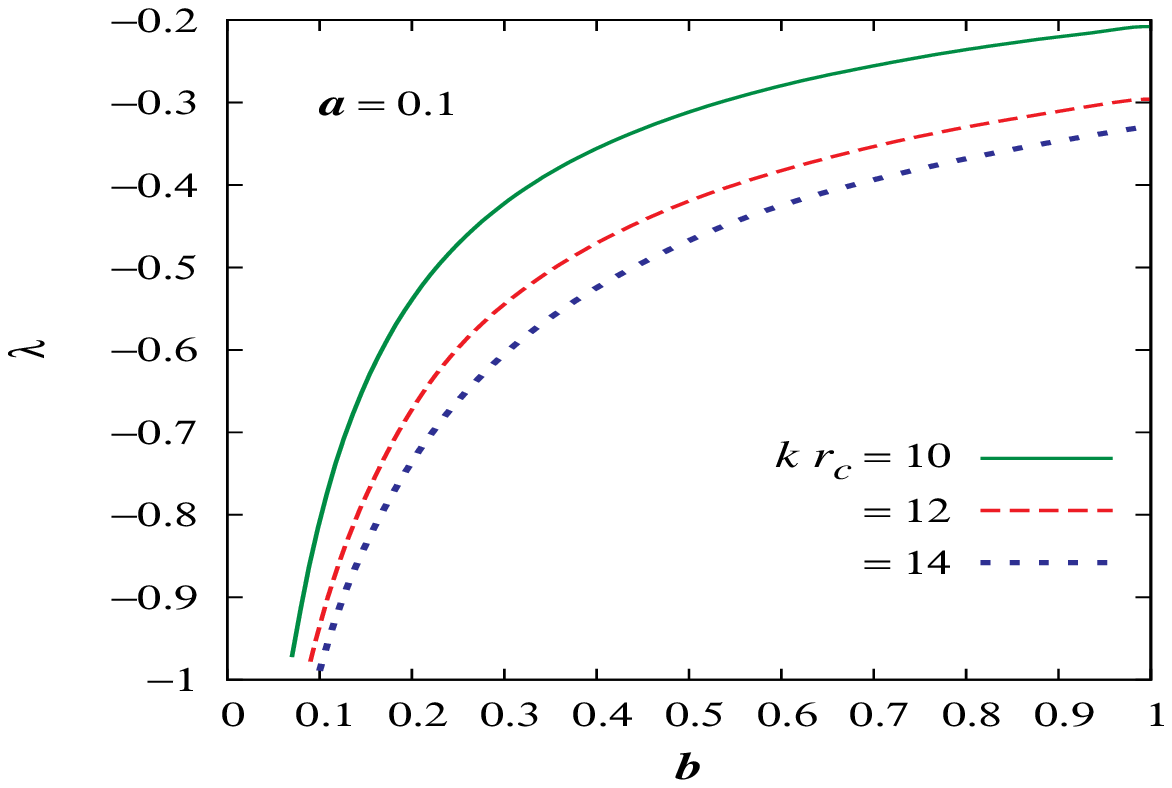}
\end{center}
\vspace*{-4ex}
\caption{$k\, r_c$ contour in the 
$(b, \lambda)$ plane for $a = 0$ (left panel) and $a = 0.1$ (right panel).}
\label{fig:krc_contour_al}
\end{figure}
\begin{figure}
\begin{center}
\includegraphics[width=2.6in,height=1.5in,angle=0]{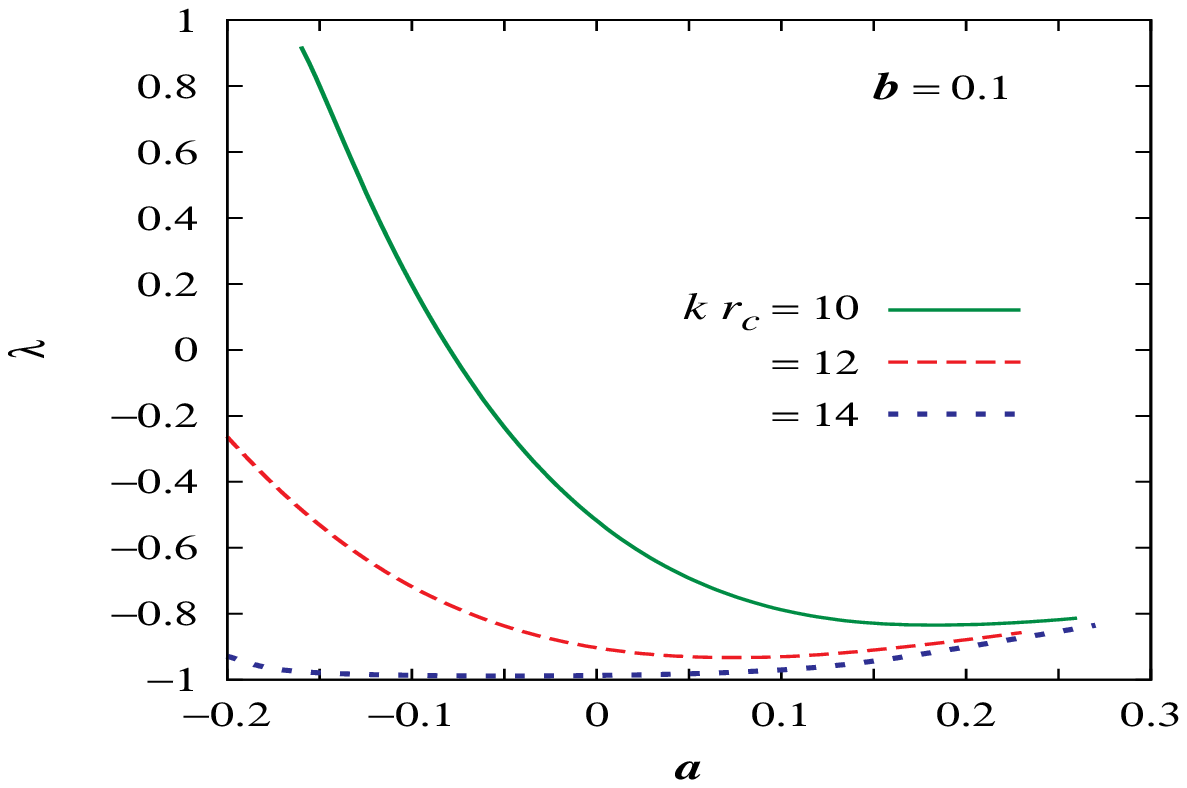}
\includegraphics[width=2.6in,height=1.5in,angle=0]{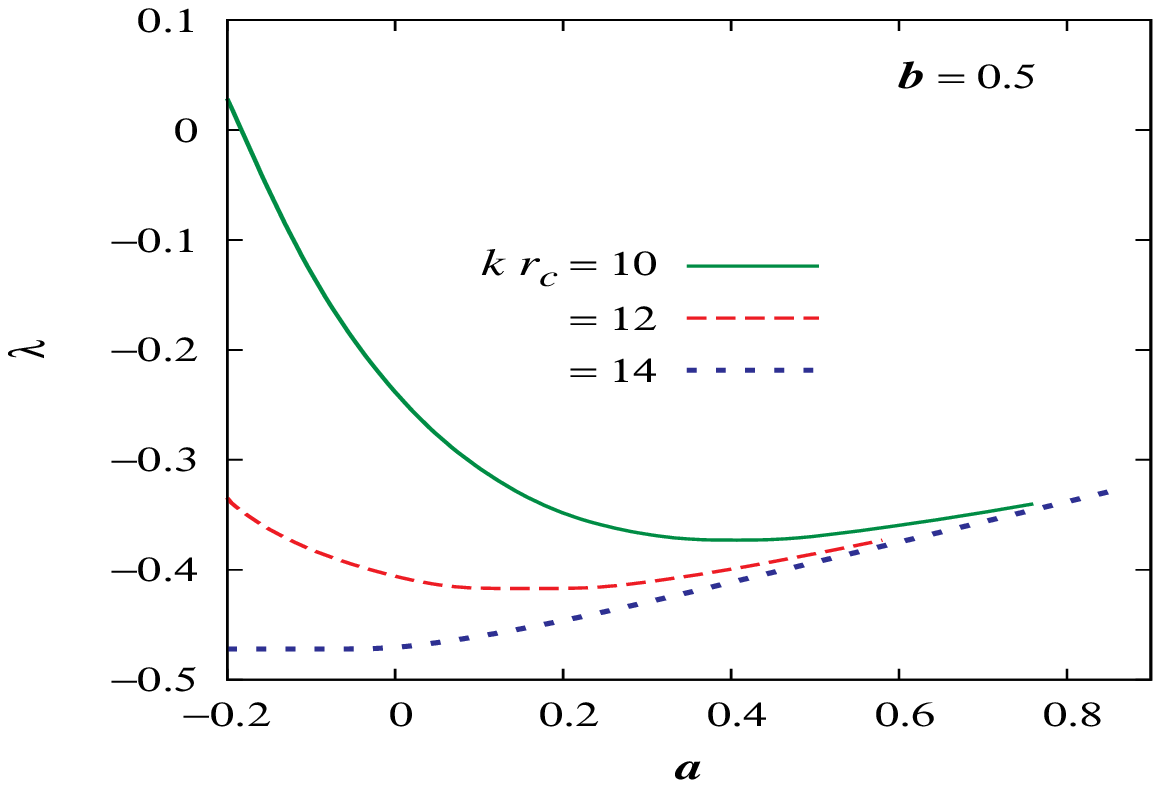}
\end{center}
\vspace*{-4ex}
\caption{$k\, r_c$ contour in the 
$(a, \lambda)$ plane for $b = 0.1$ (left panel) and $b = 0.5$ (right panel).}
\label{fig:krc_contour_b}
\end{figure}

%%%%%%%%%%%%%%%%%%%%%%%%%%%%%%%%%%%%%%
Approximating $\nu\simeq 2+\epsilon$ with $\epsilon = V_2/2k^2$, 
and denoting 
\[
n_0= 1 - \, \frac{17}{8} \,
  \sqrt{\frac{25\kappa_0\kappa_2^3-\kappa_2^2 -768}
             {(\kappa_0 \, \kappa_2 -4) \, 
              (1008 + 250\sqrt{3}\kappa_2+25\kappa_2^2)}
       } 
\]
one obtains 
the minimum (corresponding to the negative sign above) at 
\beq
kl \simeq  \, -\epsilon^{-1} \, \ln n_0 \ .
\label{min}
\eeq
%%%%%%%%%%%%%%%%%%%%%
giving us the interrelationship between the parameters
that would lead to a desired hierarchy (see
Fig.\ref{fig:krc_contour_al}). To understand the figure, note that for
$V_{1} \rightarrow 0$, $n_0 \rightarrow 0.0726 $ leading to a direct
correspondence between the hierarchy and $V_2 /2k^2$.  Since $V_2$ has
only a very weak dependence on $b$, this implies $\lambda \propto
1/\sqrt{b}$ for a given $k l$, a relation exhibited to a very large
degree by the curves in Fig.\ref{fig:krc_contour_al}. This clearly
rules out the possibility of $b=0$ and indicates the importance of
$R^3$ term.  On the other hand, $a = 0$ is clearly admissible. 
While the relationship between $a$ and $\lambda$ 
(Fig.\ref{fig:krc_contour_b}) is more complicated, it is interesting to 
note that the iso-hierarchy curves tend to a fixed point in this plane
with the location depending on the value of $b$. 

It is important to note that, unlike in the case of the
original GW mechanism~\cite{gw}, we do not have the liberty of
choosing an `appropriate' ratio of the values of the scalar field on
the two branes. Since our scalar field $\phi$ is of a geometrical
origin, we are not even allowed to introduce brane-localized
potentials unless accompanied by a corresponding change in the
geometrodynamics. In other words, the values of $\phi$ at the two
branes are fixed by the warp factor $\sigma(\yt)$. This, in turn, fixes 
the value of the brane tensions $\lambda_{v,h}$ which are no longer 
equal and opposite, but differ slightly in magnitude (to the same extent 
as $\sigma(\yt)$ differs from linearity) so that the entire solution 
is a self-consistent one. As mentioned in the beginning, this is but a 
consequence of incorporating the back-reaction (in the Einstein frame)
or, equivalently, the non-linear form of $f(\tilde R)$ (in the Jordan frame).

%%%%%%%%%%%%%%%%%% FIGURE %%%%%%%%%%%

As it turns out, the hierarchy contour is a fairly sensitive function
of the $f(R)$ parameters. This, per se, is not a negative feature of
the scenario, for the exact magnitude of the warping is unknown and
depends on the exact value of the Higgs mass on the Planck
brane. Indeed, there is a strong disagreement between the Kaluza-Klein
(KK)-graviton masses in the theory derived under the assumption that
the bulk curvature be sufficiently small compared to the fundamental
mass scale $M$ so as to permit a classical
treatment~\cite{Davoudiasl:1999jd} and the experimental bounds on the
same~\cite{ATLAS:2011ab}. It has been argued~\cite{Das:2013lqa} that
this disagreement can be alleviated if there exists a cutoff in the
theory approximately two orders of magnitude lower than $M$ with the
(unknown) physics intervening between this cutoff and $M$ responsible
for the remaining small hierarchy. While the prescription of
ref.\cite{Das:2013lqa} is an {\em ad hoc} one, the scenario we discuss
here provides a concrete threshold in the shape of $m$, the mass of
the scalar. With the latter being of geometrodynamical origin, its
value is determined the same quantum corrections that determine the
hierarchy and, indeed, there is a nonzero correlation between the
two. For example, if one were to start with a six-dimensional
doubly-warped scenario~\cite{Choudhury:2006nj} (which has been shown
to evade this tension~\cite{Arun:2014dga}), a non-trivial $f(\tilde
R)$ would be generated, if one were to integrate out one of the warped
directions.

To summarise, we have shown that the modulus field in the RS scenario
can be stabilized in purely geometrical way. Appealing to plausible
quantum corrections to the Einstein-Hilbert action, we trade the
higher derivatives of the metric tensor for an equivalent scalar field
with a complicated potential form and a nonminimal coupling to
gravity.  On going over to the Einstein frame (characterized by a
nonminimal coupling), the corresponding potential is seen to have a
local minimum leading to an negative effective bulk cosmological
constant, and a fluctuation field with a naturally small mass. The
resulting framework leads to the stabilization of the modulus without
the need to appeal to boundary localized interactions or neglecting
the backreaction. The correct hierarchy is obtained for a wide range
of parameters. Moreover, the mechanism offers a natural way out of the
tension between the theoretical expectations for the KK-graviton
masses and the strong bounds obtained at the LHC.

\bib{99}
\bibitem{cms-higgs}
%\cite{Chatrchyan:2012ufa}
%\bibitem{Chatrchyan:2012ufa} 
  S.~Chatrchyan {\it et al.}  [CMS Collaboration],
  %``Observation of a new boson at a mass of 125 GeV with the CMS experiment 
  %at the LHC,''
  Phys.\ Lett.\ B {\bf 716}, 30 (2012)
  [arXiv:1207.7235 [hep-ex]].
  %%CITATION = ARXIV:1207.7235;%%
  
\bibitem{atlas-higgs}
%\cite{Aad:2012tfa}
%\bibitem{Aad:2012tfa} 
  G.~Aad {\it et al.}  [ATLAS Collaboration],
  %``Observation of a new particle in the search for the Standard Model Higgs 
  %boson with the ATLAS detector at the LHC,''
  Phys.\ Lett.\ B {\bf 716}, 1 (2012)
  [arXiv:1207.7214 [hep-ex]];\\
  %%CITATION = ARXIV:1207.7214;%%  
%\cite{Aad:2012an}
%\bibitem{Aad:2012an} 
  G.~Aad {\it et al.}  [ATLAS Collaboration],
  %``Combined search for the Standard Model Higgs boson in $pp$ collisions at 
  %$\sqrt{s} = 7$ TeV with the ATLAS detector,''
  Phys.\ Rev.\ D {\bf 86}, 032003 (2012)
  [arXiv:1207.0319 [hep-ex]].
  %%CITATION = ARXIV:1207.0319;%%

\bibitem{rs}
%\cite{Randall:1999ee}
 % \bibitem{Randall:1999ee} 
  L.~Randall and R.~Sundrum,
  %``A Large mass hierarchy from a small extra dimension,''
  Phys.\ Rev.\ Lett.\  {\bf 83}, 3370 (1999)
  [hep-ph/9905221];\\
  %%CITATION = HEP-PH/9905221;%%
  %6067 citations counted in INSPIRE as of 02 Feb 2014
%
%\cite{Randall:1999vf}
% \bibitem{Randall:1999vf} 
%  L.~Randall and R.~Sundrum,
  %``An Alternative to compactification,''
  Phys.\ Rev.\ Lett.\  {\bf 83}, 4690 (1999)
  [hep-th/9906064].
  %%CITATION = HEP-TH/9906064;%%
  %4972 citations counted in INSPIRE as of 02 Feb 2014

\bibitem{gw} W.~D.~Goldberger and M.~B.~Wise Phys. Rev. Lett. {\bf 83} 4922 (1999).

%\cite{Cline:2000xn}
\bibitem{Cline:2000xn} 
  J.~M.~Cline and H.~Firouzjahi,
  %``Brane world cosmology of modulus stabilization with a bulk scalar field,''
  Phys.\ Rev.\ D {\bf 64}, 023505 (2001)
  [hep-ph/0005235];\\
  %%CITATION = HEP-PH/0005235;%%
  %40 citations counted in INSPIRE as of 26 Feb 2014
%\cite{Choudhury:2000wc}
%\bibitem{Choudhury:2000wc} 
  D.~Choudhury, D.~P.~Jatkar, U.~Mahanta and S.~Sur,
  %``On stability of the three 3-brane model,''
  JHEP {\bf 0009}, 021 (2000)
  [hep-ph/0004233].
  %%CITATION = HEP-PH/0004233;%%
  %11 citations counted in INSPIRE as of 26 Feb 2014

\bibitem{weyl} 
H.~ Weyl, Sitzungsber.Preuss.Akad.Wiss.Berlin (Math.Phys.)
{\bf 1918}, 465 (1918).\\
R.~H.~ Dicke, Phys.Rev.{\bf 125}, 2163–2167 (1962).

\bibitem{felice} A.~De~ Felice and S. Tsujikawa, Living Rev.Rel. {\bf 13} 3 (2010).
%\cite{Nojiri:2010wj}
\bibitem{Nojiri:2010wj} 
  S.~Nojiri and S.~D.~Odintsov,
  %``Unified cosmic history in modified gravity: from F(R) theory to Lorentz non-invariant models,''
  Phys.\ Rept.\  {\bf 505}, 59 (2011)
  [arXiv:1011.0544 [gr-qc]].
  %%CITATION = ARXIV:1011.0544;%%
  %740 citations counted in INSPIRE as of 18 Nov 2014

\bibitem{back_reaction}
%\cite{DeWolfe:1999cp}
%\bibitem{DeWolfe:1999cp} 
  O.~DeWolfe, D.~Z.~Freedman, S.~S.~Gubser and A.~Karch,
  %``Modeling the fifth-dimension with scalars and gravity,''
  Phys.\ Rev.\ D {\bf 62}, 046008 (2000)
  [hep-th/9909134];\\
  %%CITATION = HEP-TH/9909134;%%
  %725 citations counted in INSPIRE as of 02 Feb 2014
%\cite{Csaki:2000zn}
%\bibitem{Csaki:2000zn} 
  C.~Csaki, M.~L.~Graesser and G.~D.~Kribs,
  %``Radion dynamics and electroweak physics,''
  Phys.\ Rev.\ D {\bf 63}, 065002 (2001)
  [hep-th/0008151];\\
  %%CITATION = HEP-TH/0008151;%%
  %278 citations counted in INSPIRE as of 02 Feb 2014
%
%\cite{Csaki:1999mp}
%\bibitem{Csaki:1999mp} 
  C.~Csaki, M.~Graesser, L.~Randall and J.~Terning,
  %``Cosmology of brane models with radion stabilization,''
  Phys.\ Rev.\ D {\bf 62}, 045015 (2000)
  [hep-ph/9911406];\\
  %%CITATION = HEP-PH/9911406;%%
  %476 citations counted in INSPIRE as of 02 Feb 2014
%
%\cite{Maity:2006ub}
%\bibitem{Maity:2006ub} 
  D.~Maity, S.~SenGupta and S.~Sur,
  %``Stability analysis of the Randall-Sundrum braneworld in presence of bulk scalar,''
  Phys.\ Lett.\ B {\bf 643}, 348 (2006)
  [hep-th/0604195].
  %%CITATION = HEP-TH/0604195;%%
  %21 citations counted in INSPIRE as of 02 Feb 2014

%\cite{Davoudiasl:1999jd}
\bibitem{Davoudiasl:1999jd} 
  H.~Davoudiasl, J.~L.~Hewett and T.~G.~Rizzo,
  %``Phenomenology of the Randall-Sundrum Gauge Hierarchy Model,''
  Phys.\ Rev.\ Lett.\  {\bf 84}, 2080 (2000)
  [hep-ph/9909255].
  %%CITATION = HEP-PH/9909255;%%
  %374 citations counted in INSPIRE as of 25 Feb 2014

%\cite{ATLAS:2011ab}
\bibitem{ATLAS:2011ab} 
  G.~Aad {\it et al.}  [ATLAS Collaboration],
  %``Search for Extra Dimensions using diphoton events in 7 TeV proton-proton collisions with the ATLAS detector,''
  Phys.\ Lett.\ B {\bf 710}, 538 (2012)
  [arXiv:1112.2194 [hep-ex]].
  %%CITATION = ARXIV:1112.2194;%%
  %49 citations counted in INSPIRE as of 25 Feb 2014

%\cite{Das:2013lqa}
\bibitem{Das:2013lqa} 
  A.~Das and S.~SenGupta,
  %``126 GeV Higgs and ATLAS bound on the lightest graviton mass in Randall-Sundrum model,''
  arXiv:1303.2512 [hep-ph].
  %%CITATION = ARXIV:1303.2512;%%
  %2 citations counted in INSPIRE as of 25 Feb 2014
  
  %\cite{Choudhury:2006nj}
\bibitem{Choudhury:2006nj} 
  D.~Choudhury and S.~SenGupta,
  %``Living on the edge in a spacetime with multiple warping,''
  Phys.\ Rev.\ D {\bf 76}, 064030 (2007)
  [hep-th/0612246].
  %%CITATION = HEP-TH/0612246;%%
  
 %\cite{Arun:2014dga}
\bibitem{Arun:2014dga} 
  M.~T.~Arun, D.~Choudhury, A.~Das and S.~SenGupta,
  %``Graviton modes in multiply warped geometry,''
  arXiv:1410.5591 [hep-ph].
\ebib
\end{document}